\documentclass[%
 reprint,
superscriptaddress,
amsthm,amsmath,amssymb,
 aps,
]{revtex4-1}

\usepackage{graphicx}
\usepackage{dcolumn}
\usepackage{bm}
\usepackage{physics}
\usepackage{mathrsfs}
\usepackage{amssymb}
\usepackage{subfigure}  
 \usepackage{booktabs}
 \usepackage{array}

\begin{document}

\title{Quantum Random Number Generation with Partial Source Assumptions  }

\author{Xing Lin}
\email{xingl@hku.hk}
\affiliation{Department of Physics, University of Hong Kong, Pokfulam Road, Hong Kong}
\author{Rong Wang}
\email{rwangphy@hku.hk}
\affiliation{Department of Physics, University of Hong Kong, Pokfulam Road, Hong Kong}

\begin{abstract}
	Quantum random number generator harnesses the power of quantum mechanics to generate true random numbers, making it valuable for various scientific applications. However, real-world devices often suffer from imperfections that can undermine the integrity and privacy of generated randomness. To combat this issue, we present a novel quantum random number generator and experimentally demonstrate it. Our approach circumvents the need for exhaustive characterization of measurement devices, even in the presence of a quantum side channel. Additionally, we also do not require detailed characterization of the source, relying instead on reasonable assumptions about encoding dimension and noise constraints. Leveraging commercially available all-fiber devices, we achieve a randomness generation rate of 40 kbps.
\end{abstract}

\pacs{Valid PACS appear here}
\maketitle

\section{\label{sec:level1}INTRODUCTION}

Randomness is an important resource in many fields, and generating randomness that satisfies statistical properties and privacy requirements is a crucial problem. Pseudo or classical random number generators rely on determined algorithms or physical processes, which makes them vulnerable to outside attackers with enough computing power \cite{heninger2012mining}, limiting their application in privacy-sensitive areas like cryptography. Quantum random number generators (QRNG) exploit the intrinsic randomness of quantum mechanics \cite{born1926quantenmechanik}, making them a promising solution. QRNG has been extensively researched based on various models for ideal, well-characterized devices from trusted manufacturers \cite{jennewein2000fast,stipvcevic2007quantum,wahl2011ultrafast,gabriel2010generator,qi2010high,guo2010truly,bustard2011quantum,ma2016quantum,herrero2017quantum,hu2020quantum,argillander2022tunable}. However, practical devices are often complex or untrusted, and device characterization is usually incomplete and asynchronous with randomness generation, providing side channels for attackers to predict generated bits.

To address the device problem, a device-independent (DI) QRNG is a feasible solution \cite{pironio2010random,christensen2013detection,acin2016certified,bierhorst2018experimentally,liu2018device,liu2021device}. DI-QRNG utilizes the correlations observed when measuring entangled particles, allowing for the existence of both classical and quantum side channels in devices. However, the practicality of DI-QRNG is challenging due to the high demand setup of the loophole-free violation of Bell test and the low generation rate. Semi-device-independent (Semi-DI) QRNGs have been proposed as an alternative solution, with a fast rate and low demand for setup at the cost of limiting the partial power of attackers.

Many research studies on Semi-DI QRNGs have focused on untrusted randomness sources, as seen in \cite{cao2016source,marangon2017source,avesani2018source,drahi2020certified,michel2019real,zheng2020bias,fiorentino2007secure,lin2022certified,avesani2022unbounded}, which aim to develop source-independent QRNGs that can resist side channels in the source. However, in practical experiments, the measurement devices used are also complex and prone to imperfections \cite{li2019quantum,lin2020security,ma2020practical}. To address this issue, considering the classical side channels, some researchers have provided analytical randomness bounds with dimension limitations \cite{lunghi2015self,lin2022certified}, while others have focused on fully characterizing the source \cite{cao2015loss,nie2016experimental}. Additionally, numerical methods have been employed to analyze uncharacterized measurements \cite{brask2017megahertz,tebyanian2021semi,bischof2017measurement}. Recently, researchers have extended the attacks of the measurement to the quantum attacks with full characterization of the source \cite{wang2023provably}. Nevertheless, current protocols with fewer assumptions in the measurement typically require full characterization of the source to ensure the privacy of the random numbers.

In this paper, we propose a novel Semi-DI QRNG protocol and experimentally demonstrate its feasibility. Our protocol allows us to bypass the characterization of the arbitrary countable-dimensional measurement with the presence of a quantum side channel. In particular, we also do not need a detailed characterization of the source part, and only require some assumptions regarding the encoding dimension and noise constraints. One key idea in our protocol is that no measurement device can accurately forge the observable expectations of a set of indistinguishable states. By using a combination of test states, even if they are imperfect, we can provide an analytical bound on the extractable randomness solely based on the observable expectations without the need for detailed characterization of the devices. Furthermore, we demonstrate a proof-of-principle experiment using an all-fiber implementation system with a coherent source. Despite the imperfections in the modulation and detection devices, we achieve a randomness generation rate of 40 kbps.

\section{\label{sec:2}PROTOCOL DESCRIPTION}

\begin{figure}[b]
	\centering
	\includegraphics[width=0.8\linewidth]{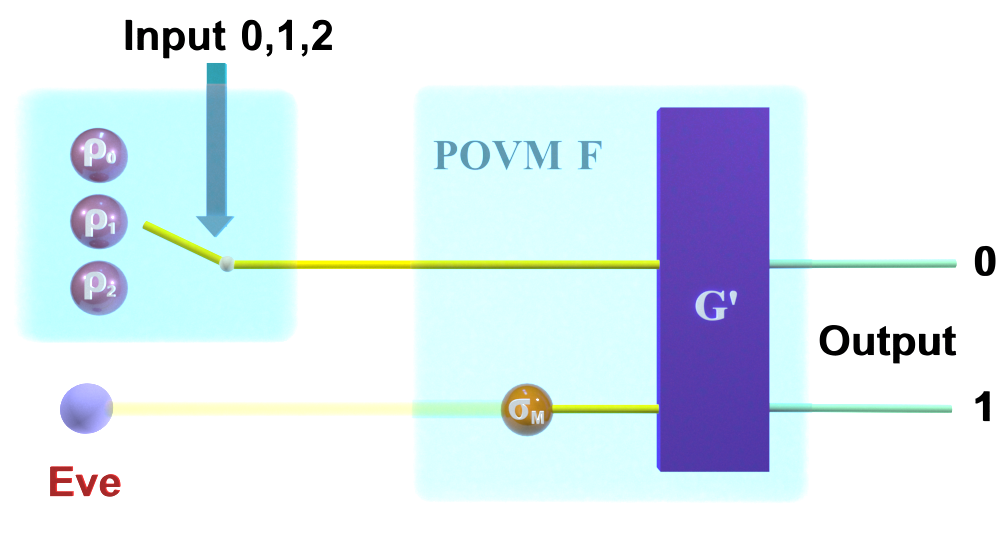}
	\caption[]{The structure of our work. The source and measurement are not fully characterized, and there may be a quantum side channel present in the measurement part. The POVM can be viewed as a projective measurement $\textbf{G'}$ that measures both the source state and the ancillary state $\sigma_{M}$}
	\label{fig:usm}
\end{figure}

The main structure of our protocol is illustrated in Fig. \ref{fig:usm}. Our protocol follows a prepare-and-measurement setup. In the source part, the protocol executor, Alice, randomly selects one qubit state from the set $\{\rho_{0}, \rho_{1}, \rho_{2}\}$ as the input state. The choice of the state is based on the corresponding input random bit $x_i$ which can take values $0$, $1$, or $2$ with unbalanced probabilities $P_g+P_t$, $P_t$, and $P_t$ respectively. These states may have imperfections and noise, but ideally, they correspond to $\{\ketbra{0},~\ketbra{+},~\ketbra{-}\}$, where $\ketbra{0}$ is one of the eigenstates of the Pauli matrices $\sigma_{z}$, and the rest of states are the eigenstates of $\sigma_{x}$. 

In the measurement part, the input state is measured by an uncharacterized countable dimensional positive operator-valued measure (POVM) $F$, resulting in a binary outcome $b_i$ that can take values $0$ or $1$. Eve may be the producer of the measurement devices and can preshare the entanglement in the measurement. Ideally, the measurement corresponds to a projective measurement of the $X=\{\ketbra{+},\ketbra{-}\}$ basis. In the post-processing part, Alice estimates the parameter using the outputs corresponding to the three states in the test rounds and then bounds the randomness generation rate $l$. The detailed protocol steps are listed in Table 1.

\begin{table}[htbp]
	\noindent
	\begin{tabular}{>{\raggedright\arraybackslash}m{8.4cm}}
		\textbf{Table.1} Protocol steps \\ \hline \\ 
		$ \bullet $ \textbf{Source:} In each of the $N$ experimental rounds, Alice randomly selects one state from $\{\rho_{0}, ~\rho_{1}, ~\rho_{2}\}$ as the input state based on the corresponding input random bit $x_i=0, 1, 2$, with probabilities $P_g+P_t$, $P_t$ and $P_t$, respectively. Ideally, the three states respectively correspond to $\{\ketbra{0},~\ketbra{+},~\ketbra{-}\}$.\\
		$ \bullet $ \textbf{Measurement:} The input state is measured in each round by the uncharacterized POVM $F$, resulting in a binary outcome $b_i=0$ or $1$. Ideally, the measurement is the $X=\{\ketbra{+},\ketbra{-}\}$ basis.\\
		$ \bullet $ \textbf{Randomness generation:} After completing the rounds, Alice selects $NP_g$ binary outcome bits from the settings where $x_i=0$ to obtain the raw random sequence. \\
		$ \bullet $ \textbf{Parameter estimation:} Using the remaining $3NP_t$ outcome bits, Alice can bound the parameter $C$ and estimate the randomness generation rate $l$. If the estimation of $C$ fails or $l$ is negative, the rounds are aborted. \\
		$ \bullet $ \textbf{Randomness extraction:}  Alice applies a $universal_2$ hash function to extract $l$ final random bits from the raw sequence. The security of the protocol is guaranteed by the composable security and quantum leftover hashing lemma, with a security parameter $\varepsilon_{t}$. \\
		\hline
	\end{tabular}
	\label{tab:tab1}
\end{table}

Our protocol is based on several key assumptions. (\romannumeral1)-The protocol consists of a trusted but error-prone source and an uncharacterized measurement which may have a quantum side channel. (\romannumeral2)-The states of the source are two dimensional. (\romannumeral3)-The purity of the generation state $\rho_{0}$ is higher than that of the test states $\rho_{1}$ and $\rho_{2}$. (\romannumeral4)-The system is subject to an independent and identically distributed (i.i.d.) process.

Assumption (\romannumeral1) forms the fundamental structure of our protocol. We allow the presence of some imperfections in both source and measurement components of the protocol, which may be known to Eve. Furthermore, we allow Eve to preshare the entanglement in the measurement. However, we must assume that the measurement devices cannot transmit information to the outside world during the execution of the protocol. The assumption regarding the untrusted measurement has also been addressed in a previous QRNG study \cite{wang2023provably}, distinguishing it from the assumption made in measurement-device-independent quantum key distribution \cite{lo2012measurement}. Assumptions (\romannumeral2) and (\romannumeral3) impose limitations on the state preparation. Assumption (\romannumeral2) is one of the conditions to ensure the indistinguishability of the generation state and the test states. To fulfill (\romannumeral2), it is necessary that the effective light pulse contains no more than one photon. Here we simulate the behavior of a single photon source using a phase-randomized coherent source by estimating the proportion of single photon and vacuum, while ensuring that the encoding space remains independent of the photon number space. Assumption (\romannumeral3) limits the amount of noise or contamination in the states. In the case of qubit states, we can equivalently express the purity relation as the generation state $\rho_{0}$ having a longer Bloch vector compared to the test states $\rho_{1}$ and $\rho_{2}$ \cite{gamel2012measures}. To fulfill (\romannumeral3), it is necessary to have a lower modulation noise corresponding to the generation state $\rho_{0}$. Assumption (\romannumeral4) implies that our protocol is designed to defend against collective attacks on the measurement. In the supplementary materials, we will provide a detailed discussion on the assumptions satisfied by our implementation.

\section{\label{sec:3}Security framework}

We now present our main result, and the detailed proof can be found in the supplementary materials. The objective of our protocol is to estimate genuine randomness by measuring the expectations of three input qubit states. Since the output is limited to a binary outcome, we can prove that any countable-dimensional POVM can be represented as a two-dimensional POVM $\textbf{F}$ with two elements, $\{F_{0}, F_{1}\}$. In this context, we define the input states $\rho_{0}$, $\rho_{1}$, and $\rho_{2}$ corresponding to the vectors $\vec{S}_{0}$, $\vec{S}_{1}$, and $\vec{S}_{2}$ on the Bloch sphere. And regarding to the measurement, we represent the elements $F_{0}=a_{0}I_{2}+\frac{\vec{T}}{2}\cdot\vec{\sigma}$ and $F_{1}=(1-a_{0})I_{2}-\frac{\vec{T}}{2}\cdot\vec{\sigma}$ with the Bloch vector $ \vec{T}$ \cite{lunghi2015self,cao2015loss} and the two-dimensional identity matrix $I_{2}$. To establish a bound of the extractable randomness from $\rho_{0}$, we define the parameter $C$ as

\begin{eqnarray}
	C=\abs{\vec{T}\times\vec{S}_{0}}	.
\end{eqnarray}
Here, $C$ implies the extractable randomness, with a maximum value of 1 indicating the highest randomness scenario. To establish a lower bound for $C$, we consider the observable expectations of the input states $\rho_{0}$, $\rho_{1}$, and $\rho_{2}$, denoted as $g_{0}$, $g_{1}$, and $g_{2}$ respectively. These observable expectations are defined as $g_{i}=\Tr[(F_{0}-F_{1})\rho_{i}]~(i=1,2,3)$. Based on the geometric properties of Bloch vectors, we can derive a lower bound on $C$ using these observable expectations by

\begin{eqnarray}
	C\geq\sqrt{(g_{1}-g_{0})(g_{0}-g_{2})}.
\end{eqnarray}
Note that this result implies that during the parameter estimation step, we should retain the experimental results that satisfy $(g_{1}-g_{0})(g_{0}-g_{2})\geq0$ and abort the protocol if this condition is not met.

To bound the extractable randomness by $C$, we need to estimate the guessing probability $p^{g}_{guess}(A|\rho_{0}, \textbf{F})$ with the generation state $\rho_{0}$ and the POVM $\textbf{F}$. In the case of a classical side channel on the state, we assume a decomposition of the state $\rho_{0}=\sum_{j}q_{j}\ketbra{\omega_{j}}$. According to the Naimark theorem, we assume that Eve has access to the purification $\ket{\psi_{ME}}$ of the ancillary state $\sigma_{M}=\Tr_{E}[\ketbra{\psi_{ME}}]$ in the measurement. The measurement is performed using a projective measurement $\textbf{G'}=\{G'_{1},\dots G'_{n}\}$ that measures both the source state $\rho_{0}$ and the ancillary state $\sigma_{M}$ as shown in Fig. \ref{fig:usm}. To distinguish the different outputs, Eve uses the measurement $M^{E}_{k}$ to measure her parts of the purification. In this case, by combining the duality idea for each pure state $\ket{\omega_{j}}$ to a projective measurement \cite{cao2015loss}, and considering the concavity of the guessing probability, we can derive the upper bound of the guessing probability $p^{q}_{guess}(A| \rho_{0}, \textbf{F}) $ by

\begin{eqnarray}
	p^{q}_{guess}(A|\rho_{0}, \textbf{F})=&&\max_{\{\{G'_{k}\}_{k},\{M^{E}_{k}\}_{k},q_{j},\ket{\omega_{j},\psi_{ME}}\}}\sum_{j,k}q_{j}\nonumber\\&&\Tr[G'_{k}\otimes M^{E}_{k}\ketbra{\omega_{j},\psi_{ME}}]\nonumber\\\leq&&1-\frac{C}{2}\left( 1-\sqrt{1-C^{2}}\right).
\end{eqnarray}

Finally, we calculate the length of the final randomness bit with using a phase-randomized coherent source to simulate the single photon source. Considering the statistical fluctuations of $C$ estimation and photon number in the coherent source for the finite data, we can use the quantum leftover hash lemma \cite{tomamichel2011leftover} to establish a lower bound on the length of final randomness by

\begin{eqnarray}\label{}
	&&l\geq- N_{g}(\eta+\theta_{g})\log_{2}\left(1-\frac{C}{2}\left( 1-\sqrt{1-C^{2}}\right)\right)\nonumber\\&&\qquad-2\log_{2}\frac{1}{2\varepsilon},\nonumber\\
	&&C\geq\frac{1}{\eta}\sqrt{\left( g_{e1}-g_{e0}-2( 1-\eta+2\theta_{t})\right) \left( g_{e0}-g_{e2}\right) },\qquad
\end{eqnarray}
where $\theta_{t}=\sqrt{\ln(1/\varepsilon)/(2N_{t})}$ and $\theta_{g}=\sqrt{\ln(1/\varepsilon)/(2N_{g})}$ are the statistic fluctuation parameter, with $\varepsilon$ being the failure probability. $N_{g}$ and $N_{t}$ represent the number of generation rounds and test rounds. $\eta=(1+\mu)/e^{\mu}$ denotes the probability of the photon number being no larger than 1 with an average photon number of $\mu$. $g_{e0}$, $g_{e1}$, and $g_{e2}$ represent the experimental results corresponding to $g_{0}$, $g_{1}$, and $g_{2}$, respectively. With the consideration of composable security, the total failure probability satisfies $\varepsilon_{t}=7\varepsilon$.

We note that as our protocol is designed for randomness expansion, it only needs an initial true random seed. Unlike self-testing QRNG protocols that require additional secure pseudo-random numbers to test the devices \cite{lunghi2015self,brask2017megahertz,tebyanian2021semi,lin2022certified}, our protocol does not have this requirement. This means that the presence of an eavesdropper, who could potentially access the pseudo-random numbers, is not a concern. Consequently, our protocol is more secure and better suited to withstand outside attacks.

\section{\label{sec:4}EXPERIMENT}

\begin{figure}
	\centering
	\includegraphics[width=1\linewidth]{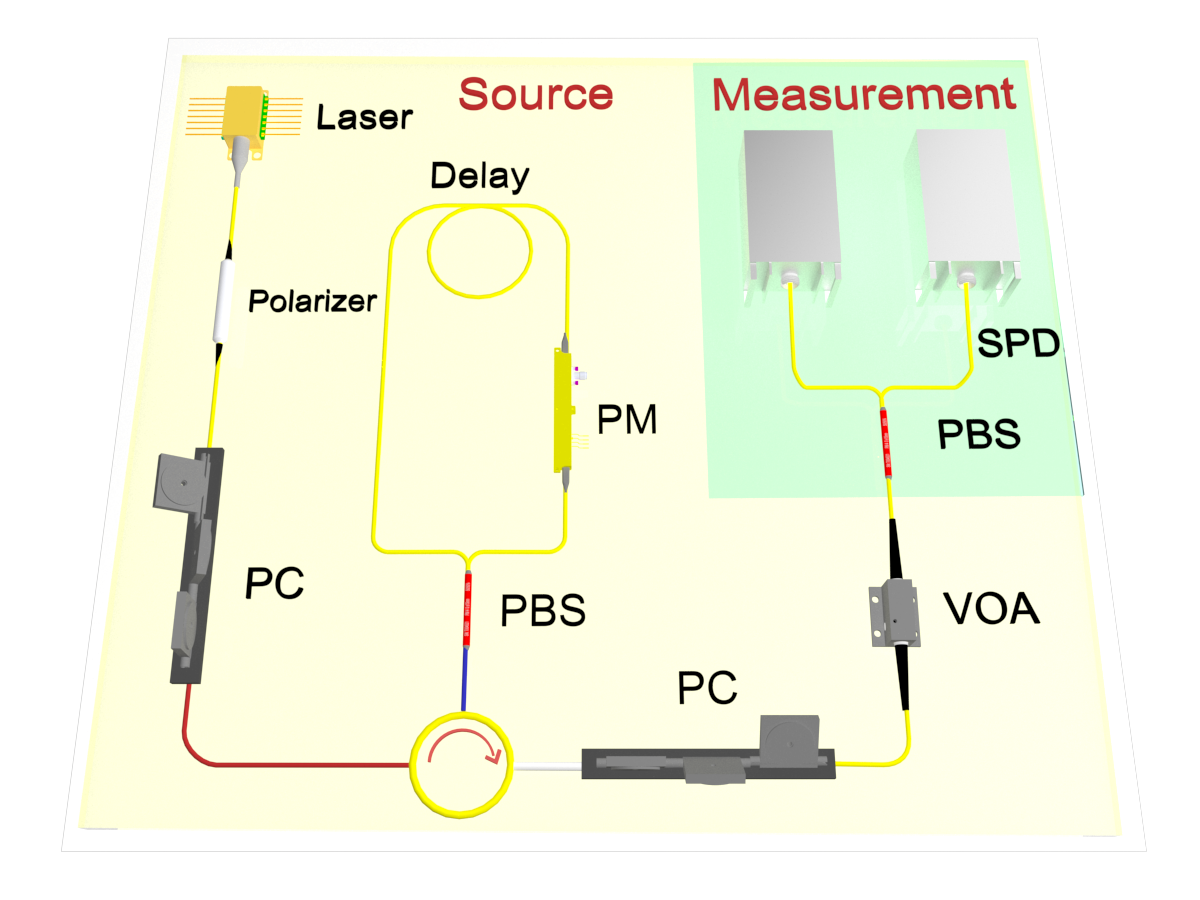}
	\caption[]{Experiment setup for the protocol. We uses a pulse laser as a phase-randomized coherent source, which is modulated by two polarization controllers (PC) and a Sagnac loop with a phase modulator (PM) to choose different states. A polarization beam splitter (PBS) and two single photon detectors (SPD) are used as a measurement $\{\ketbra{H},~\ketbra{V}\}$. PC, polarization controller; PBS, polarization beam splitter; PM, phase modulator; VOA, variable optical attenuator; SPD, single photon detectors.}
	\label{fig:experiment}
\end{figure}

To show the feasibility of the protocol, we set up an all-fiber proof-of-principle experiment system with the polarization encoding method, as displayed in Fig.\ref{fig:experiment}. Our protocol does not require precise preparation of the state and measurement. However, to achieve a high performance in terms of the randomness rate, precise modulation of the state and measurement is beneficial.

We utilize a 10 MHz gain-switched pulse laser (Eblana Photonics EP1550-NLW-B) as a phase-randomized coherent light source. To achieve a $\ket{\phi}=\ket{H}+\exp(i\varphi)\ket{V}$ polarization state, we modulate the output photon in each pulse using a fiber polarizer and a polarization controller (PC, Thorlabs FPC562). We then input this state into a Sagnac loop that consists of a polarization beam splitter (PBS, Thorlabs PBC1550SM-APC), a phase modulator (PM, iXblue MPZ-LN-10), and a 3 m fiber delay.

By using an arbitrary waveform generator (Siglent SDG6052X), we introduce a random signal to modulate the PM with $\theta_{c}$ and $\theta_{a}$ (clockwise and anticlockwise) phase modulation. For all states, we set $\theta_{c}=0$. For the $\rho_{0}$ state, we choose $\theta_{a0}=0$, and for the $\rho_{1}$ and $\rho_{2}$ state, we choose $\theta_{a1}=\frac{\pi}{2}$ and $\theta_{a2}=-\frac{\pi}{2}$, respectively. Due to practical modulation error of the PM, there is a total extra misalignment error \cite{fan2019modeling} of $\Delta\theta_{m}=\frac{\pi}{14}$ rad for the $\rho_{1}$ and $\rho_{2}$ states, which satisfies $\Delta\theta_{m}=\pi-\theta_{a1}+\theta_{a2}$.

The output states from the Sagnac loop are then modulated by a second polarization controller (PC) to rotate $\rho_{0}$, $\rho_{1}$ and $\rho_{2}$ from the polarizations $\ket{H}+\exp(i(\theta_{ai}+\varphi))\ket{V}~(i=1,2,3)$ to the polarizations $\ket{H}+\exp(i\varphi')\ket{V}$, $\ket{H}$ and $\ket{V}$, respectively. Finally, we adjust the loss using a variable optical attenuator (VOA, Thorlabs EVOA1550A) to generate the output states.

For the measurement, we use a PBS and two single photon detectors (SPD, ID Qube NIR Gated) as a measurement $\{\ketbra{H},~\ketbra{V}\}$, with the SPDs in gated mode with 10 MHz, 3 ns gates. The detection efficiencies of two SPDs are 10.6\% and 13.7\%, and the dark count probabilities are $1.3\times 10^{-6}$ and $1.6\times 10^{-6}$. We use a time-digital converter (ID1000 Time Controller) to collect the response signals and assign the click of detector $H$ as 0 and the click of detector $V$ as 1. The no-click and double-click events will be assigned a value of 0.

\begin{figure}
	\centering
	\includegraphics[width=0.95\linewidth]{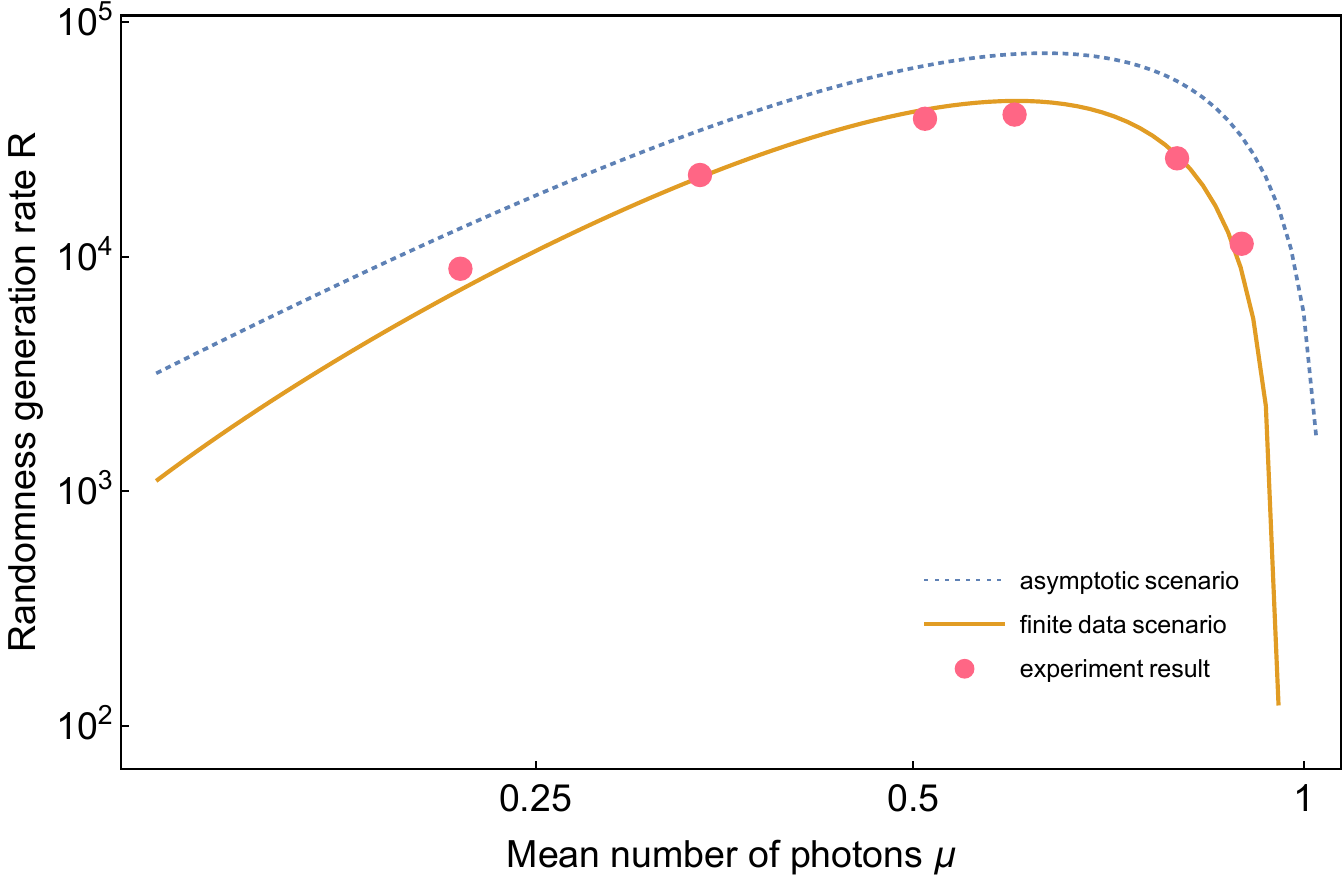}
	\caption[]{The results of the randomness generation rate from the experiment, as well as from the simulation of asymptotic and finite data scenarios. The system frequency was set at 10 MHz and the round number of a block was $N=10^{10}$, with a total failure probability of $\varepsilon_{t}=7\times10^{-10}$. The mean photon numbers selected for the experiment were 0.21, 0.33, 0.49, 0.58, 0.78, and 0.89.}
	\label{fig:simulation}
\end{figure}

\begin{figure}
	\centering
	\includegraphics[width=0.95\linewidth]{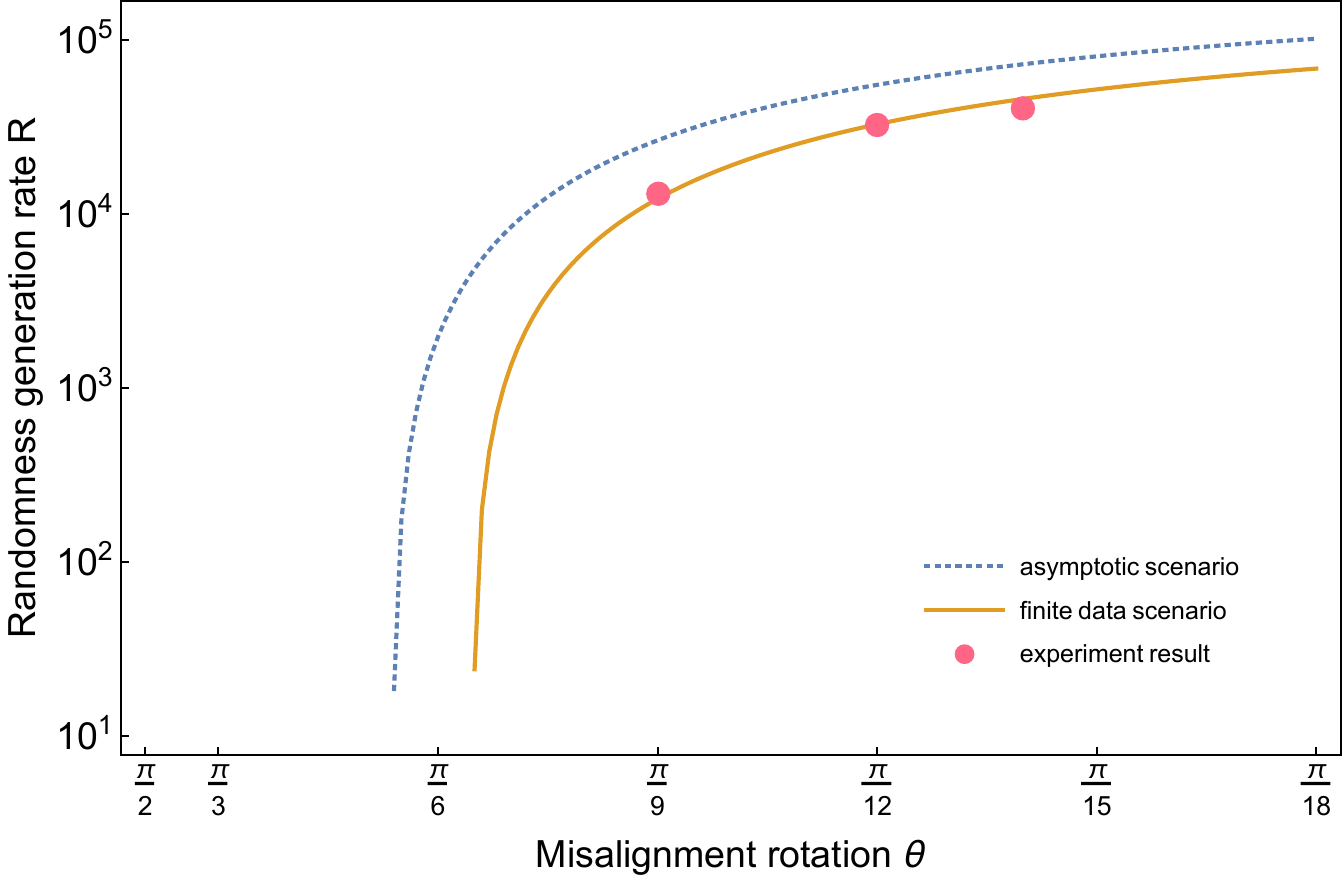}
	\caption[]{The results of the randomness generation rate from the experiment of different modulation error, as well as from the simulation. Here we choose the optimal intensity of 0.58 photon per pulse. The modulation errors of the rotations for experiment are $ \frac{\pi}{9}$, $ \frac{\pi}{12}$, $ \frac{\pi}{14}$.}
	\label{fig:rotation}
\end{figure}

In Fig. \ref{fig:simulation}, we present the experimental results with different intensities, as well as the simulation results for both the asymptotic and finite data cases. The mean photon numbers of per pulse after the total loss chosen as 0.21, 0.33, 0.49, 0.58, 0.78, and 0.89. Here as detection efficiency mismatch, we choose the common loss contribution of the detectors is 0.106. We choose $N=10^{10}$ bits as a block to estimate the randomness rate of each intensity, with a total failure probability of $\varepsilon_{t}=7\times10^{-10}$. The maximum rate achieved was 40.415 kbps in the experiment with a mean photon number $\mu$ of 0.58 per pulse, corresponding to $\sim$0.004 bit per pulse. We observed that as the mean photon numbers $\mu$ increased, the randomness generation rate also increased due to fewer no-click events lacking randomness. However, when $\mu$ was larger than 0.58, the rate quickly decreased due to a higher proportion of multiphoton events lacking randomness. When $\mu$ approaches 1, the rate became 0. 

In Fig. \ref{fig:rotation}, we present the simulation results and the experiment results with different modulation error $ \frac{\pi}{9}$, $ \frac{\pi}{12}$, $ \frac{\pi}{14}$, using the optimal mean photon numbers 0.58. The results indicate that a slight error does not noticeably affect the randomness rate, demonstrating the robustness of the protocol towards imperfections. However, when the error reaches $\frac{\pi}{9}$, the randomness rate decreases rapidly and reaches 0 at around $\frac{\pi}{6}$ error.

Finally, the private random numbers are extracted by the Toeplitz-matrix hashing. The final random bits successfully passed all the tests in the NIST test suite \cite{rukhin2001statistical}. The detailed data have been shown in the supplementary materials. 

\section{\label{sec:6}discussion}

In this work, we propose a Semi-DI QRNG that does not require a detailed characterization of both the source and measurement, and allows for the presence of a quantum side channel in the measurement. By analyzing the observable expectations of the test states, we can synchronously monitor the min-entropy of the raw data. We implement our protocol using an all-fiber experimental system with a coherent source, and achieve a rate of over 40 kbps. Compared to previous Semi-DI QRNG protocols that aimed to address imperfect measurement \cite{cao2015loss,nie2016experimental,brask2017megahertz,tebyanian2021semi,bischof2017measurement,wang2023provably}, our QRNG offers a method with an analytical bound, further reducing the characterization required in the devices without significantly sacrificing practicality of the protocol.

Our protocol and proof-of-principle experiment can be improved in several ways. Firstly, incorporating a high-frequency detector \cite{zhang201916} and a high-rate single photon source \cite{ma2020ultrabright} could directly enhance the randomness generation rate to tens of Mbps in our implementation. This improvement would directly contribute to the overall effectiveness of our protocol. Additionally, the removal of the i.i.d. assumption in our protocol will expose it to both coherent attacks and collective attacks, thereby expanding the potential attack abilities of Eve. This is an important improvement that requires further research. Various methods, such as entropy accumulation theory \cite{metger2022generalised} or numerical analysis \cite{zhou2023numerical}, are currently being explored to address this challenge. We are optimistic that these approaches can also be effectively applied to enhance the security of our protocol. Our protocol is one of the efforts to further relax device assumptions without compromising practicality, making QRNGs more practical in various applications.

\begin{acknowledgments}
	We thank Hoi-Kwong Lo for inspirational discussions and valuable comments. We also thank Wenyuan Wang and Chengqiu Hu for helpful discussions. This work was supported by the University of Hong Kong start-up grant. X. L. also acknowledged support from the Research Grants Council of Hong Kong (AoE/P-701/20).
	\hfill
	
	X. L. and R.W. contributed equally to this work.
\end{acknowledgments}

\nocite{*}
\bibliography{apssamp}

\appendix

\onecolumngrid

\section*{Supplementary materials}

\subsection{Guessing probability estimation}

In this section, we will provide a detailed proof of the estimation of the guessing probability. Firstly, we will reduce the problem of arbitrary measurement to the two-dimensional measurement case. Then, we will derive the bound of $C$ in this scenario. Next, we will provide an upper bound of the guessing probability for the pure states input and extend it to the case of mixed states, connecting it with $C$. Finally, we will extend the analysis from classical attacks to quantum attacks.

\noindent \textit{Step 1: reducing arbitrary measurement to the two dimensional measurement with two elements}

Our protocol involves the input states represented by the qubit states and an output limited to two values with eigenvalues $\pm1$. By considering the Naimark extension, any POVM $\textbf{M}=\{M_{1},\dots,M_{n}\}$ can be seen as an extended project-value measurement (PVM) $\textbf{G}=\{G_{1},\dots G_{n}\}$ and a large unitary operator $ U_{AM}$ with an ancilla $\sigma_{M}$, as illustrated in Fig. \ref{fig:usm}. We can combine the PVM and the unitary operator to get a new PVM $\textbf{G'}=\{G'_{1},\dots G'_{n}\}$ by \cite{dai2022intrinsic} 

\begin{eqnarray}
	G'_{i}=U_{AM}^{\dagger}G_{i}U_{AM}.
\end{eqnarray}

Considering a decomposition of the ancillary state $\sigma_{M}=\sum_{j}\lambda_{j}\ketbra{\tau_{Mj}}$, we can provide a corresponding decomposition of the POVM $\textbf{M}=\sum_{j}\lambda_{j}\textbf{N}_{j}$, where each POVM $\textbf{N}_{j}=\{N_{j1},\dots,N_{jn}\}$. This allows us to represent the probability of obtaining measurement result $k$ for the input state $\rho_i~(i=0,1,2)$ by

\begin{eqnarray}
	\Tr[\rho_{i}M_{k}]&=&\Tr[\rho_{i}\left( \sum_{j}\lambda_{j}N_{jk}\right) ]\nonumber\\&=&\Tr[\sum_{j}\lambda_{j}(\rho_{i}\otimes\ketbra{\tau_{Mj}})G'_{k}]\nonumber\\&=&\Tr_{E}\left[ \rho_{i}\left( \sum_{j}\lambda_{j}\Tr_{A}\left[ \left( I_{A}\otimes\ketbra{\tau_{Mj}}\right)G'_{k}\right] \right) \right] ,
\end{eqnarray}
where $I_{A} $ is the two-dimensional identity matrix on the state space. We can define the POVM $\textbf{M'}=\{M'_{1},\dots,M'_{n}\}$, where the element $M'_{k} $ satisfies

\begin{eqnarray}
	M'_{k}=\sum_{j}\lambda_{j}\Tr_{M}[\left( I_{A}\otimes\ketbra{\tau_{Mj}}\right)G'_{k}],
\end{eqnarray}
As we can observe, the partial trace operator removes the ancillary space $\mathbb{M}$ in each element $M'_{k}$. This allows each $M'_{k}$ to operate solely within this two-dimensional state space. Consequently, we can consider the process of the POVM $\textbf{M}$ operating on the qubit state $\rho_i$ as an equivalent process of the two-dimensional POVM $\textbf{M'}$ operating on the state $\rho_i$.

In the scenario where there are only two outputs, we can group the $n$ elements $\{M'_{1},\dots,M'_{n}\}$ into two elements $\mathbf{F}=\{F_{0},F_{1}\}$. Here, $F_{0}=\sum_{i} M^{'0}_{i}$ and $F_{1}=\sum_{i} M^{'1}_{i}$. The elements $M^{'0}_{i}$ and $M^{'1}_{i}$ correspond to the components that produce outputs 0 and 1, respectively, from the set $\{M'_{1},\dots,M'_{n}\}$. Therefore, we can represent the POVM, regardless of its dimension, by the two-dimensional POVM $\textbf{F}=\{F_{0},F_{1}\}$, which can be decomposed using the Pauli matrices, such as

\begin{eqnarray}
	F_{0}&=&a_{0}I_{2}+\frac{\vec{T}}{2}\cdot\vec{\sigma}\nonumber\\
	F_{1}&=&(1-a_{0})I_{2}-\frac{\vec{T}}{2}\cdot\vec{\sigma},
\end{eqnarray}
where $I_{2}$ is the 2 dimensional identity matrix, and $ \vec{T}=(T_{x}, T_{y}, T_{z})$ is a vector in the Bloch sphere. $a_{0}$ is the parameter corresponding to classical imperfections, which satisfies $a_{0}\in[0, 1]$. We define the POVM operator $F=F_{0}-F_{1}=(2a_{0}-1)I_{2}+\vec{T}\cdot\vec{\sigma}$.

\hfill\\

\noindent \textit{Step 2: bounding $C$ by the observable expectations}

\begin{figure}
	\centering
	\includegraphics[width=0.4\linewidth]{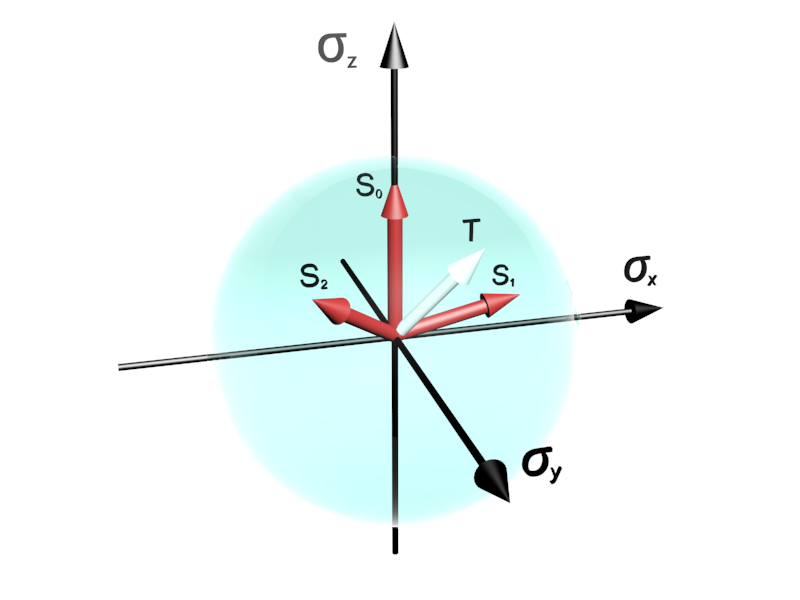}
	\caption[]{A schematic diagram of the Bloch vectors in the Bloch sphere. $\vec{S}_{0}$, $\vec{S}_{1}$, and $\vec{S}_{2}$ represent the Bloch vectors of the states $\rho_{0}$, $\rho_{1}$ and $\rho_{2}$, respectively. The length of $\vec{S}_{0}$ is larger than that of $\vec{S}_{1}$ and $\vec{S}_{2}$. Meanwhile, $\vec{T}$ represents the Bloch vector of the POVM $\textbf{F}=\{F_{0},F_{1}\}$.}
	\label{fig:xzp}
\end{figure}

In the following, we will define and bound $C$ using the observable expectations. Let three two-dimensional states $\rho_{0}$, $\rho_{1}$, and $\rho_{2}$ correspond to the vectors $\vec{S}_{0}$, $\vec{S}_{1}$, and $\vec{S}_{2}$ in the Bloch sphere, respectively. In Fig. \ref{fig:xzp}, we demonstrate the vectors $\vec{S}_{0}$, $\vec{S}_{1}$, $\vec{S}_{2}$, and $\vec{T}$ in the Bloch sphere. Without loss of generality, we set the vector $\vec{S}_{0}$ on the z-axis in the figure.

As defined in the main text, we define

\begin{eqnarray}\label{6}
	C=\abs{\vec{T}\times\vec{S}_{0}}=\sqrt{\abs{\vec{T}}^{2}\abs{\vec{S}_{0}}^{2}-(\vec{T}\cdot\vec{S}_{0})^{2}}.	
\end{eqnarray}
In the definition of $C$, we can observe that $C$ reaches its maximum value of 1 if and only if $\abs{\vec{T}}=\abs{\vec{S}_{0}}=1$ and $\vec{T}$ and $\vec{S}_{0}$ are orthogonal. This scenario represents the highest randomness generation, with 1 bit of true randomness being generated each round. On the other hand, when $C=0$, it means that either $\abs{\vec{T}}=0$, or $\abs{\vec{S}_{0}}=0$, or $\vec{T}$ and $\vec{S}_{0}$ are parallel. In this case, it is obvious that we cannot generate randomness with this combination of the state and measurement. Therefore, we can infer that $C$ is a parameter connected to the extractable randomness. However, we cannot obtain the value of $C$ just from the measurement results of the state $\rho_{0}$. Hence, in the following, we will attempt to bound $C$ combining the measurement results of the introduced two test states $\rho_{1}$ and $\rho_{2}$.

As our assumption $\abs{\vec{S}_{0}}\geq\abs{\vec{S}_{1}}$ and $\abs{\vec{S}_{0}}\geq\abs{\vec{S}_{2}}$, we can get $\abs{\vec{T}}\abs{\vec{S}_{0}}\geq\vec{T}\cdot\vec{S}_{1}$ and $\abs{\vec{T}}\abs{\vec{S}_{0}}\geq-\vec{T}\cdot\vec{S}_{2}$. Then we can get

\begin{eqnarray}\label{7}
	C&=&\sqrt{(\abs{\vec{T}}\abs{\vec{S}_{0}}-\vec{T}\cdot\vec{S}_{0})(\abs{\vec{T}}\abs{\vec{S}_{0}}+\vec{T}\cdot\vec{S}_{0})}\nonumber\\&\geq&\sqrt{(\vec{T}\cdot\vec{S}_{1}-\vec{T}\cdot\vec{S}_{0})(-\vec{T}\cdot\vec{S}_{2}+\vec{T}\cdot\vec{S}_{0})}.
\end{eqnarray}

Based on the definitions of observable expectations in the main text, we can connect the observable expectations of the states $\rho_{0}$, $\rho_{1}$, and $\rho_{2}$ to their Bloch vectors. Specifically, we have $g_{0}=\Tr[F\rho_{0}]=(2a_{0}-1)+\vec{T}\cdot\vec{S}_{0}$, $g_{1}=\Tr[F\rho_{1}]=(2a_{0}-1)+\vec{T}\cdot\vec{S}_{1}$, and $g_{2}=\Tr[F\rho_{2}]=(2a_{0}-1)+\vec{T}\cdot\vec{S}_{2}$. Then, based on the derivations above, we can obtain the lower bound of $C$ by the observable expectations, which satisfies

\begin{eqnarray}\label{}
	C&\geq&\sqrt{(\vec{T}\cdot\vec{S}_{1}-\vec{T}\cdot\vec{S}_{0})(-\vec{T}\cdot\vec{S}_{2}+\vec{T}\cdot\vec{S}_{0})}\nonumber\\&=&\sqrt{(2a_{0}-1+\vec{T}\cdot\vec{S}_{1}-(2a_{0}-1+\vec{T}\cdot\vec{S}_{0}))(-(2a_{0}-1+\vec{T}\cdot\vec{S}_{2})+2a_{0}-1+\vec{T}\cdot\vec{S}_{0})}\nonumber\\&=&\sqrt{(g_{1}-g_{0})(g_{0}-g_{2})}.
\end{eqnarray}
In the derivations above, we need $(\vec{T}\cdot\vec{S}_{1}-\vec{T}\cdot\vec{S}_{0})(-\vec{T}\cdot\vec{S}_{2}+\vec{T}\cdot\vec{S}_{0})\geq0$. We can remain the results satisfied $(g_{1}-g_{0})(g_{0}-g_{2})\geq 0 $ and abort the protocol if it doesn't satisfy. Since the observable expectations are limited to the range of $[-1,1]$, we can see that this bound will achieve its maximum value if and only if $g_{1}-g_{0}=g_{0}-g_{2}=1$ (considering that $g_{1}\geq g_{0}\geq g_{2}$). This corresponds to the scenario where the generation state is $\ket{0}$, and the test states are $\ket{+}$ and $\ket{-}$, which are non-orthogonal with the generation state. In this case, a projective measurement $\{\ketbra{+},\ketbra{-}\}$ is used. We can observe that since these states are indistinguishable, any eavesdropper cannot correctly falsify the observable expectations corresponding to the maximum value of $C$ by presetting the measurement. Thus, it corresponds to the situation of private randomness generation.

Here, we have connected the observable expectations $g_{0}$, $g_{1}$, and $g_{2}$ with the parameter $C$. However, we cannot obtain the extractable randomness only from the current form of $C$. Next, we will consider how to bound the guessing probability by using $C$.

\hfill\\

\noindent \textit{Step 3: bounding $p_{guess}$ of the pure state}

We now consider the possibility of a classical eavesdropper for the measurement. To carry out a classical attack on the measurement, Eve can preset the form of the decomposition of the projective measurements $\textbf{P}_i$. The POVM $\textbf{F}$ can always be decomposed into a sum of two-dimensional extremal POVMs which consists the projective measurement and the trivial measurement $\{I_{2},0\}$ \cite{d2005classical,cao2015loss}, that is

\begin{eqnarray}
	F_{0}&=&a_{0}I_{2}+\frac{\vec{T}}{2}\cdot\vec{\sigma}=\sum_{i}p_{i}P_{i0} \nonumber\\
	F_{1}&=&(1-a_{0})I_{2}-\frac{\vec{T}}{2}\cdot\vec{\sigma}=\sum_{i}p_{i}P_{i1}+(1-2a_{0})I_{2}\nonumber\\
	\sum_{i}&p_{i}&=2a_{0},
\end{eqnarray}
where $\textbf{P}_{i}=\{P_{i0},P_{i1}\}$ is the two-dimensional projective measurement (if $a_{0}>0.5$, we can use $1-a_{0}$ to replace $a_{0}$ without loss of generality).

Here we first consider the situation of the pure state $\ket{\omega}$ as the input generation state. In this case, we can establish a duality between any POVM with pure state and the corresponding mixed state with projective measurement, which is similar to the idea presented in ref. \cite{cao2015loss}. For each projective measurement $\{P_{i0}=\ketbra{\psi_{i0}},P_{i1}=\ketbra{\psi_{i1}}\}$ and pure input state $\ket{\omega}$, the guessing probability $p_{guess}(A|\ketbra{\omega}, \textbf{P}_{i}) $ only depends on the inner product of $\max\{\abs{\bra{\psi_{i0}}\ket{\omega}},\abs{\bra{\psi_{i1}}\ket{\omega}}\}$. This means that we can add a unitary operator $u_{i}$ to establish the duality between the pure state $\ket{\omega}$ and each projective measurement $\textbf{P}_{i} $. Note that in this context, we define $u_{i}$ to be a rotation operator around the axis which is orthogonal to the plane supported by the Bloch vectors of $\ket{\omega}$ and $\ket{\psi_{i0}}$ in the Bloch sphere. This rotation is performed in an anticlockwise direction, with an angle no more than $\pi$.

As a result, we let $\ket{\psi_{i0}}=u_{i}\ket{\omega} $ and $\ket{\psi_{i1}}=u_{i}\ket{\omega_{\perp}} $, $\ket{\omega} $ and $\ket{\omega_{\perp}} $ are the pure states which have opposite Bloch vectors. The guessing probability $p_{guess}(A|\ketbra{\omega}, \textbf{F}) $ can be expressed as: 

\begin{eqnarray}\label{a8}
	p_{guess}(A|\ketbra{\omega}, \textbf{F})&=&\max_{\{p_{i},\textbf{P}_{i}\}}\sum_{i}p_{i}\max_{k=0,1}\Tr[P_{ik}\ketbra{\omega}]+(1-2a_{0})\Tr[I_{2}\ketbra{\omega}]\nonumber\\
	&=&\max_{\{p_{i},\textbf{P}_{i}\}}\sum_{i}p_{i}\max\{\abs{\bra{\psi_{i0}}\ket{\omega}},\abs{\bra{\psi_{i1}}\ket{\omega}}\}+(1-2a_{0})\nonumber\\
	&=&\max_{\{p_{i},u^{\dagger}_{i}\}}\sum_{i}p_{i}\max\{\abs{\bra{\omega}u^{\dagger}_{i}\ket{\omega}},\abs{\bra{\omega_{\perp}}u^{\dagger}_{i}\ket{\omega}}\}+(1-2a_{0})\nonumber\\
	&=&2a_{0}\max_{\{p_{i},\ket{\psi'_{i}}\}}\sum_{i}\frac{p_{i}}{2a_{0}}\max\{\abs{\bra{\omega}\ket{\psi'_{i}}}^{2},\abs{\bra{\omega_{\perp}}\ket{\psi'_{i}}}^{2}\}+(1-2a_{0}),
\end{eqnarray}
where $\ket{\psi'_{i}}=u^{\dagger}_{i}\ket{\omega} $. In Eq. \ref{a8}, we can explain the first term by considering the noise source scenario that the dual input qubit state $\rho_{F}=\sum_{i}\frac{p_{i}}{2a_{0}}\ketbra{\psi'_{i}}$ is measured by the projective measurement $\textbf{w}=\{\ketbra{\omega},\ketbra{\omega_{\perp}}\}$. We let $p_{guess}(A|\rho_{F},\textbf{w}) $ as the guessing probability of this situation, we can get

\begin{eqnarray}\label{a9}
	p_{guess}(A|\ketbra{\omega}, \textbf{F})&=&2a_{0}\max_{\{p_{i},\ket{\psi'_{i}}\}}\sum_{i}\frac{p_{i}}{2a_{0}}\max\{\abs{\bra{\omega}\ket{\psi'_{i}}}^{2},\abs{\bra{\omega_{\perp}}\ket{\psi'_{i}}}^{2}\}+(1-2a_{0})\nonumber\\
	&=&2a_{0}p_{guess}(A|\rho_{F},\textbf{w})+(1-2a_{0}).
\end{eqnarray}

The guessing probability with the noise source and the projective measurement has been widely researched in source-independent QRNG \cite{fiorentino2007secure} and coherence of formation\cite{yuan2015intrinsic,cao2015loss,dai2022intrinsic}. For the qubit state $\rho_{F}$ and the projective measurement $\textbf{w}$, we can get

\begin{eqnarray}\label{a10}
	p_{guess}(A|\rho_{F},\textbf{w})=\frac{1+\sqrt{1-n^{2}_{xy\omega}}}{2}\qquad \abs{n_{xy\omega}}=\sqrt{n^{2}_{x\omega}+n^{2}_{y\omega}},
\end{eqnarray} 
where $n_{x\omega}=\text{Tr}[\sigma_{x\omega}\rho_F]$ and $n_{y\omega}=\text{Tr}[\sigma_{y\omega}\rho_F]$. $\sigma_{x\omega}=\ketbra{\omega}{\omega_{\perp}}+\ketbra{\omega_{\perp}}{\omega}$ and $\sigma_{y\omega}=-i\ketbra{\omega}{\omega_{\perp}}+i\ketbra{\omega_{\perp}}{\omega}$ are the Pauli matrices based on the representation of $\{\ketbra{\omega},\ketbra{\omega_{\perp}}\}$. 

Based on Eq. \ref{a9}, \ref{a10}, we can give the guessing probability $p_{guess}(A|\ketbra{\omega}, \textbf{F}) $ by

\begin{eqnarray}\label{a11}
	p_{guess}(A|\ketbra{\omega}, \textbf{F})&=&2a_{0}\frac{1+\sqrt{1-n^{2}_{xy\omega}}}{2}+(1-2a_{0})\nonumber\\
	&=&1-a_{0}\left( 1-\sqrt{1-n^{2}_{xy\omega}}\right) .
\end{eqnarray}

\hfill\\

\noindent \textit{Step 4: bounding $p_{guess}$ of the mixed state by $C$}

For the mixed state $\rho_{0}$ as the generation state, we can decompose it into the a sum of the pure state $\rho_{0}=\sum_{j}q_{j}\ketbra{\omega_{j}}$. In fact, considering the classical memory of Eve for the source, the guessing probability $p_{guess}(A| \rho_{0}, \textbf{F}) $ with mixed state $\rho_{0}$ and POVM $ \textbf{F}$ can be given by \cite{senno2022quantifying}

\begin{eqnarray}
	p_{guess}(A|\rho_{0}, \textbf{F})=\max_{\{p'_{i},\{M_{ik}\}_{k},q_{j},\ket{\omega_{j}}\}}\sum_{i,j}q_{j}p'_{i}\max_{k}\Tr[M_{ik}\ketbra{\omega_{j}}],
\end{eqnarray}
where $\{M_{ik}\}_{k}$ is a extremal decomposition of POVM $ \textbf{F}$, which satisfies $F_{k}=\sum_{i}p'_{i}M_{ik}$. Note that here we need the independence of the source and the measurement. As our assumption, the source is a trusted part and the measurement may be produced by an eavesdropper, so this requirement naturally applies to our situation. Based on the definition of guessing probability $p_{guess}(A| \rho_{0}, \textbf{F}) $, we can connect it with the guessing probability $p_{guess}(A|\ketbra{\omega}, \textbf{F}) $ with pure state and POVM in Eq. \ref{a8} by

\begin{eqnarray}
	p_{guess}(A|\rho_{0}, \textbf{F})&=&\max_{\{p_{i},\textbf{P}_{i},q_{j},\ket{\omega_{j}}\}}\sum_{i,j}q_{j}(p_{i}\max_{k=0,1}\Tr[P_{ik}\ketbra{\omega_{j}}]+(1-2a_{0})\Tr[I_{2}\ketbra{\omega_{j}}])\nonumber\\
	&\leq&\max_{\{q_{j},\ket{\omega_{j}}\}}\sum_{j}q_{j}(\max_{\{p_{i},\textbf{P}_{i}\}}\sum_{i}p_{i}\max_{k=0,1}\Tr[P_{ik}\ketbra{\omega_{j}}]+(1-2a_{0})\Tr[I_{2}\ketbra{\omega_{j}}])\nonumber\\
	&=&\max_{\{q_{j},\ket{\omega_{j}}\}}\sum_{j}q_{j}p_{guess}(A|\ket{\omega_{j}}, \textbf{F}).
\end{eqnarray}

Now we show the concavity of the guessing probability $p_{guess}(A|\ketbra{\omega_{j}}, \textbf{F}) $. We note that in Eq. \ref{a11}, $p_{guess}(A|\ketbra{\omega}, \textbf{F}) $ is a liner function of $a_{0}$, thus it is concave with respect to $a_{0}$. For $n_{xy\omega}\in[-1,1] $, the second order derivatives of $\abs{n_{xy\omega} }$ on $p_{guess}(A|\ketbra{\omega}, \textbf{F}) $ can be given by

\begin{eqnarray}
	\frac{\partial^{2} p_{guess}(A|\ketbra{\omega}, \textbf{F})}{\partial \left( \abs{n_{xy\omega}}\right) ^{2}}&=&\frac{\partial^{2}\left(  1-a_{0}\left( 1-\sqrt{1-n^{2}_{xy\omega}}\right)\right) }{\partial \left( \abs{n_{xy\omega}}\right) ^{2}}\nonumber\\
	&=&\frac{-a_{0}}{\sqrt{1-n^{2}_{xy\omega}}^{3}}\leq0.
\end{eqnarray}
Since the second order derivatives is negative, the concavity holds for $\abs{n_{xy\omega}} $. As the concavity of the guessing probability $p_{guess}(A|\ketbra{\omega}, \textbf{F}) $, we can get the upper bound of $p_{guess}(A| \rho_{0}, \textbf{F}) $ by

\begin{eqnarray}
	p_{guess}(A|\rho_{0}, \textbf{F})&\leq&\max_{\{q_{j},\ket{\omega_{j}}\}}\sum_{j}q_{j}p_{guess}(A|\ket{\omega_{j}}, \textbf{F})\nonumber\\
	&=&\max_{\{q_{j},\ket{\omega_{j}}\}}\sum_{j}q_{j}\left( 1-a_{0}\left( 1-\sqrt{1-n^{2}_{xy\omega j}}\right)\right)  \nonumber\\
	&\leq&\max_{\{q_{j},\ket{\omega_{j}}\}}\left( 1-a_{0}\left( 1-\sqrt{1-(\sum_{j}q_{j}\abs{n_{xy\omega j}})^{2}}\right)\right) .
\end{eqnarray}
Similarly, here $\abs{n_{xy\omega j}}=\sqrt{n^{2}_{x\omega j}+n^{2}_{y\omega j}} $ is the parameter corresponding to the state $\ket{\omega_{j}} $, where $n_{x\omega j}=\text{Tr}[\sigma_{x\omega j}\rho_{Fj}]$ and $n_{y\omega j}=\text{Tr}[\sigma_{y\omega j}\rho_{Fj}]$. $\sigma_{x\omega j}=\ketbra{\omega_{j}}{\omega_{j\perp}}+\ketbra{\omega_{j\perp}}{\omega_{j}}$ and $\sigma_{y\omega j}=-i\ketbra{\omega_{j}}{\omega_{j\perp}}+i\ketbra{\omega_{j\perp}}{\omega_{j}}$ are the Pauli matrices based on the representation of $\{\ketbra{\omega_{j}},\ketbra{\omega_{j\perp}}\}$, and $\rho_{Fj}=\sum_{i}\frac{p_{i}}{2a_{0}}\ketbra{\psi'_{ij}}=\sum_{i}\frac{p_{i}}{2a_{0}}u^{\dagger}_{ij}\ketbra{\omega_{j}}u_{ij}$. For every $u_{ij} $, $\ket{\psi_{i0}}=u_{ij}\ket{\omega_{j}} $ and $\ket{\psi_{i1}}=u_{ij}\ket{\omega_{j\perp}} $. 

\begin{figure}
	\centering
	\includegraphics[width=0.5\linewidth]{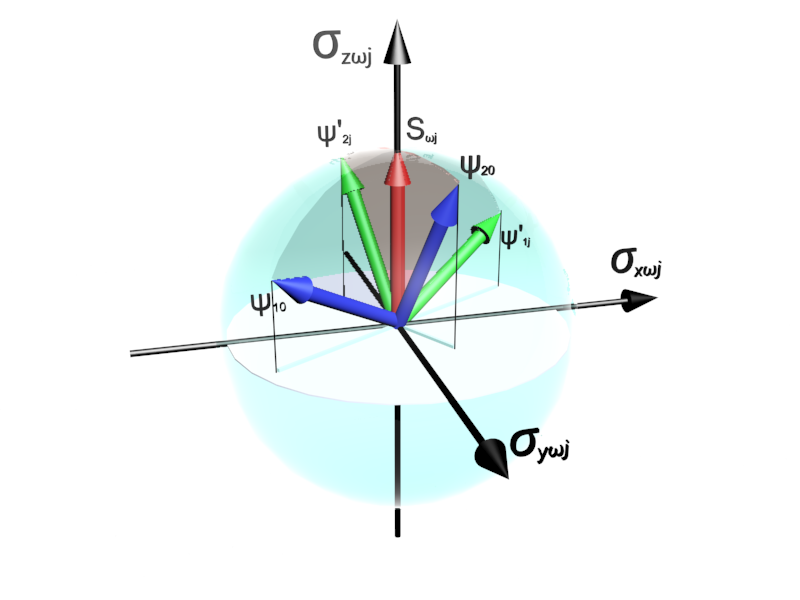}
	\caption[]{A schematic diagram of an example of the rotation operator $u_{ij}$ on the Bloch sphere, based on the representation of $\{\ketbra{\omega_{j}},\ketbra{\omega_{j\perp}}\}$. The Bloch vector of the state $\ket{\omega_{j}}$ is denoted as $\vec{S}_{\omega j}$. Since $u_{ij}$ and $u_{ij}^{\dagger}$ correspond to symmetric rotations on the same plane, the Bloch vectors of $\ket{\psi_{10}}$ and $\ket{\psi'_{1j}}$, as well as the Bloch vectors of $\ket{\psi_{20}}$ and $\ket{\psi'_{2j}}$, are symmetric around $\vec{S}_{\omega j}$. This symmetry causes the lengths of the projections of the Bloch vectors of states $\rho_{Fj}$ and $\rho'_{Fj}$ on the x-y plane to be the same.}
	\label{fig:u}
\end{figure}

We note that $\abs{n_{xy\omega j}} $ is actually the length of the projection of the Bloch vector of state $\rho_{Fj}$ on the x-y plane of the Bloch sphere based on the representation of $\{\ketbra{\omega_{j}},\ketbra{\omega_{j\perp}}\}$. When we define $ \vec{S}_{\omega j}$ as the Bloch vector of the state $\ket{\omega_{j}} $ and $ \vec{T'}_{\omega j}$ as the Bloch vector of the state $\rho_{Fj}$, we can represent $\abs{n_{xy\omega j}} $ by the cross product of the Bloch vectors by

\begin{eqnarray}
	\abs{n_{xy\omega j}}=\abs{\vec{T'}_{\omega j}\times\vec{S}_{\omega j}}.
\end{eqnarray}
Here we define the state $\rho'_{Fj}=\sum_{i}\frac{p_{i}}{2a_{0}}\ketbra{\psi_{i0}}=\sum_{i}\frac{p_{i}}{2a_{0}}u_{ij}\ketbra{\omega_{j}}u^{\dagger}_{ij}$. As per the definition, $u_{ij}$ represents an anticlockwise rotation from the Bloch vector $\vec{S}_{\omega j}$ to the Bloch vector of $\ket{\psi_{i0}}$ on the plane supported by these two vectors. Thus, $u_{ij}^{\dagger}$ correspondingly represents the clockwise rotation operator on the same plane. This implies that the Bloch vectors of $\ket{\psi_{i0}}$ and $\ket{\psi'_{ij}}$ are symmetric around $\vec{S}_{\omega j}$, as shown in the example in the Fig. \ref{fig:u}. This symmetry causes the Bloch vectors of $\rho_{Fj}$ and $\rho'_{Fj}$ to be symmetric around $\vec{S}_{\omega j}$ as well. That means the lengths of the projections of the Bloch vectors of states $\rho_{Fj}$ and $\rho'_{Fj}$ on the x-y plane, based on the representation of $\{\ketbra{\omega_{j}},\ketbra{\omega_{j\perp}}\}$, are the same.

Since $ F_{0}=a_{0}I_{2}+\frac{\vec{T}}{2}\cdot\vec{\sigma}=\sum_{i}p_{i}\ketbra{\psi_{i0}}$, we can get $\rho'_{Fj}=\sum_{i}\frac{p_{i}}{2a_{0}}\ketbra{\psi_{i0}}=\frac{1}{2}I_{2}+\frac{1}{2a_{0}}\frac{\vec{T}}{2}\cdot\vec{\sigma}$ with the Bloch vector $ \frac{\vec{T}}{2a_{0}}$. Therefore, we can get

\begin{eqnarray}
	\abs{n_{xy\omega j}}=\abs{\vec{T'}_{\omega j}\times\vec{S}_{\omega j}}=\frac{1}{2a_{0}}\abs{\vec{T}\times\vec{S}_{\omega j}}.
\end{eqnarray}
And combine with Eq. \ref{6}, we can get the lower bound of $\sum_{j}q_{j}\abs{n_{xy\omega j}} $ by

\begin{eqnarray}
	\sum_{j}q_{j}\abs{n_{xy\omega j}}=\frac{1}{2a_{0}}\sum_{j}q_{j}\abs{\vec{T}\times\vec{S}_{\omega j}}\geq\frac{1}{2a_{0}}\abs{\sum_{j}q_{j}(\vec{T}\times\vec{S}_{\omega j})}=\frac{1}{2a_{0}}\abs{\vec{T}\times\vec{S}_{0}}=\frac{C}{2a_{0}}.
\end{eqnarray}

Based on the property of $2a_{0}\in[\abs{\vec{T}},1] $ and $\abs{\vec{T}}\geq C $, we can bound $p_{guess}(A| \rho_{0}, \textbf{F}) $ according to the the monotonically decreasing of Eq. \ref{a11} by

\begin{eqnarray}
	p_{guess}(A|\rho_{0}, \textbf{F})&\leq&\max_{\{q_{j},\ket{\omega_{j}}\}}\left( 1-a_{0}\left( 1-\sqrt{1-(\sum_{j}q_{j}\abs{n_{xy\omega j}})^{2}}\right)\right) \nonumber\\
	&\leq& 1-\frac{C}{2}\left( 1-\sqrt{1-C^{2}}\right).
\end{eqnarray}

\hfill\\

\noindent \textit{Step 5: extending classical attack to quantum attack}

Here, we further extend the above result to encompass the quantum attack, which is primarily based on the result in ref. \cite{senno2022quantifying}. The above analysis focused on the scenario where the eavesdropper, Eve, has access to only the information of the mixed state and unknown measurement, and does not have access to any entanglement. This is known as the classical attack scenario. In our assumption, we consider a quantum attack scenario for the measurement, in which Eve may preshare entanglement with the ancillary state $\sigma_{M}$ in the measurement in order to obtain the maximum guessing probability for the outputs, as shown in Fig. \ref{fig:usm}. 

We assume that Eve has access to the purification $\ket{\psi_{ME}}$ of the ancillary state $\sigma_{M}=\Tr_{E}[\ket{\psi_{ME}}]$ in the measurement. According to the Naimark theorem we discussed above, the measurement is performed using a PVM $\textbf{G'}=\{G'_{1},\dots G'_{n}\}$ that measures both the source state $\rho_{0}$ and the ancillary state $\sigma_{M}$. To distinguish the different outputs, Eve uses the measurement $M^{E}_{k}$ to measure her parts of the purification. In this case, the guessing probability $p^{q}_{guess}(A^{n}| \rho_{0}, \textbf{F}) $ with quantum attacks satisfies\cite{senno2022quantifying,avesani2022unbounded}:

\begin{eqnarray}\label{11}
	p^{q}_{guess}(A|\rho_{0}, \textbf{F})=\max_{\{\{G'_{k}\}_{k},\{M^{E}_{k}\}_{k},q_{j},\ket{\omega_{j}},\ket{\psi_{ME}}\}}\sum_{j}q_{j}\sum_{k}\Tr[G'_{k}\otimes M^{E}_{k}\ket{\omega_{j}}\ket{\psi_{ME}}\bra{\omega_{j}}\bra{\psi_{ME}}].
\end{eqnarray}

In fact, for scenarios of the quantum attack and the classical attack, the optimal parameter group $\{\{G'_{k}\}_{k},\{M^{E}_{k}\}_{k},q_{j},\ket{\omega_{j}},\ket{\psi_{ME}}\} $ is in turn one of the parameter groups $ \{p_{i},\{M_{ik}\}_{k},q_{j},\ket{\omega_{j}}\}$, and vice versa \cite{dai2022intrinsic,senno2022quantifying}. That means, considering the classical attack for the source, the guessing probability in the classical attack of measurement $p_{guess}(A|\rho_{0}, \textbf{F})$ is equal to the guessing probability in the quantum attack of measurement $p^{q}_{guess}(A|\rho_{0}, \textbf{F})$ \cite{senno2022quantifying}, thus we can get 

\begin{eqnarray}
	p^{q}_{guess}(A|\rho_{0}, \textbf{F})=p_{guess}(A|\rho_{0}, \textbf{F})\leq1-\frac{C}{2}\left( 1-\sqrt{1-C^{2}}\right).
\end{eqnarray}

Specifically, considering the case of a finite number of signals $n$ with an independent and identically distributed product state $\rho_{0}^{\otimes n}$, for the collective attack which Eve perform independent attacks to each round, the total conditional min-entropy $H_{\min}(A^{n}|E^{n})_{\rho_{0}^{\otimes n}}$ can be given by the additivity of conditional min-entropy $H_{\min}(A^{n}|E^{n})_{\rho_{0}^{\otimes n}}=nH_{\min}(A|E)_{\rho_{0}} $\cite{renner2008security}. Therefore, we can get the total guessing probability $p^{q}_{guess}(A^{n}| \rho_{0}^{\otimes n}, \textbf{F}^{\otimes n}) $ satisfies

\begin{eqnarray}\label{}
	p^{q}_{guess}(A^{n}| \rho_{0}^{\otimes n}, \textbf{F}^{\otimes n})=p^{q}_{guess}(A|\rho_{0}, \textbf{F})^{n}\leq\left( 1-\frac{C}{2}\left( 1-\sqrt{1-C^{2}}\right)\right) ^{n}.
\end{eqnarray}
Here, we note that when the value of $C$ reaches its maximum value of 1, we can obtain the upper bound $(\frac{1}{2})^{n}$ for $p^{q}_{guess}(A^{n}| \rho_{0}^{\otimes n}, \textbf{F}^{\otimes n})$, which corresponds to a min-entropy of 1 in each generation round.

Our protocol's ability to allow for quantum attacks is achieved through a combination of factors. Firstly, we allow Eve to access the purification of the ancillary state $\sigma_{M}$ in the measurement. Additionally, our protocol does not require the use of extra pseudo-random numbers to test the devices, which eliminates concerns about Eve potentially accessing those numbers. It is unlike self-testing QRNG protocols which need extra trusted pseudo-random numbers to test the devices \cite{lunghi2015self,brask2017megahertz,tebyanian2021semi}. Instead, we only require an initial true random seed, which make our protocol more secure and better suited for withstanding attacks.


\subsection{Practical source and statistical fluctuation}

In this section, we consider the effects of the parameters in the practical experiment, such as using a phase-randomized coherent source and taking into account the statistical fluctuation. Here we set the total number of rounds is denoted as $N$, which includes $N_{g}=NP_{g}$ generation rounds and $3N_{t}=3NP_{t}$ test rounds. And as our protocol, during the generation rounds, we send the states $\rho_{0}$, and during the test rounds, we choose $N_{t}$ rounds to send $\rho_{0}$, $N_{t}$ rounds to send $\rho_{1}$ and $N_{t}$ rounds to send $\rho_{2}$.

\hfill
\\
\noindent \textit{Practical source}

In our previous analysis, we made the assumption that the input state is a qubit state. However, in practical applications, commonly used light sources often contain a multiphoton component, such as coherent sources. It is evident that the presence of multiphotons will impact the indistinguishability between the generated state and the test states. To utilize these sources in practical experiments, it is necessary to eliminate the multiphoton component by estimating the proportion of single photons and vacuum. It should be noted that, in order to achieve an equivalent qubit input for a phase-randomized coherent source with a two-dimensional encoding, it is crucial that the encoding space is independent of the photon number space in practical devices. The vacuum state is considered secure and can be calculated. During measurement, the vacuum only produces predetermined clicks and does not compromise security when Eve receives the state from a phase-randomized coherent source and perceives it as a mixture of Fock states \cite{lo2005getting,cao2015loss}. To account for loss tolerance, we assign a value of 0 to no-clicks and double-clicks.

During the $N_{g}$ generation round, we calculate the single photon and vacuum components of the phase-randomized coherent source, and use this information to determine the min-entropy. Specifically, based on the connection between guessing probability and the condition min-entropy \cite{konig2009operational}, we can get:

\begin{eqnarray}\label{13}
	H_{\min}(A^{n}|E^{n})_{\rho_{0}^{\otimes n}}=-\log_{2} p^{q}_{guess}(A^{n}| \rho_{0}^{\otimes n}, \textbf{F}^{\otimes n})\geq-N_{g}\Pr[n\leq1]\log_{2}( 1-\frac{C}{2}\left( 1-\sqrt{1-C^{2}}\right)).
\end{eqnarray}
where $\Pr[n\leq1]$ is the probability of the photon number being less than 1. During the test round, we consider the worst-case scenario to estimate the value of $C$ based on the experimental results, which include $g'_{0}$, $g'_{1}$, and $g'_{2}$ (considering asymmetric situation), representing the observable expectation with the practical source in the test rounds for $\rho_{0}$, $\rho_{1}$, and $\rho_{2}$, respectively. $g'_{i}$ $ (i=0,1,2)$ is a combination of the response probabilities of single photons and vacuum components as well as multi-photon components. The observable expectation for single photons and vacuum components is given by $g_{i}$. For multi-photon components, the observable expectation can range from -1 to 1. Therefore, we can establish upper and lower bounds for $g'_{i}$ based on these probabilities

\begin{eqnarray}
	\Pr[n\leq1]g_{i}-\Pr[n>1]\leq g'_{i}\leq \Pr[n\leq1]g_{i}+\Pr[n>1].\qquad i=0,~1,~2
\end{eqnarray}
We consider the scenario where the multi-photon components cause the most significant disturbance for estimating the value of $C$. This is regarded as the worst-case scenario. Therefore, we can obtain the worst-case value of $C$ that satisfies

\begin{eqnarray}\label{18}
	C&\geq&\sqrt{(g_{1}-g_{0})(g_{0}-g_{2})}\nonumber\\&\geq&\sqrt{((g'_{1}-\Pr[n>1])\frac{1}{\Pr[n\leq1]}-g_{0})(g_{0}-(g'_{2}+\Pr[n>1])\frac{1}{\Pr[n\leq1]})}.
\end{eqnarray}
As this lower bound function is a symmetric concave function for $g_{0}$ about $g_{0}=(g'_{1}+g'_{2})/2\Pr[n\leq1] $, if we obtain the experimental result $g'_{0}\geq (g'_{1}+g'_{2})/2$, we can give the lower bound by choose $g_{0}=(g'_{0}+\Pr[n>1])\frac{1}{\Pr[n\leq1]} $, which satisfies (if $g'_{0}\leq (g'_{1}+g'_{2})/2$, we can choose $g_{0}=(g'_{0}-\Pr[n>1])\frac{1}{\Pr[n\leq1]} $.)

\begin{eqnarray}\label{}
	C\geq\frac{1}{\Pr[n\leq1]}\sqrt{(g'_{1}-g'_{0}-2\Pr[n>1])(g'_{0}-g'_{2})}.
\end{eqnarray}

Considering that in our implementation we use a phase-randomized coherent source with an average photon number of $\mu$ as input, we can obtain the min-entropy by

\begin{eqnarray}\label{16}
	&&H_{\min}(A^{n}|E^{n})_{\rho_{0}^{\otimes n}}\geq- N_{g}\eta\log_{2}\left(  1-\frac{C}{2}\left( 1-\sqrt{1-C^{2}}\right)\right) \nonumber\\
	&&C\geq\frac{1}{\eta}\sqrt{\left( g'_{1}-g'_{0}-2(1-\eta)\right) (g'_{0}-g'_{2})}.
\end{eqnarray}
where $\eta$ denotes the probability of the photon number being no larger than 1, and $\eta=(1+\mu)/e^{\mu}$ corresponding to a coherent source with an average photon number of $\mu$.

Note that decoy state method is also a way to bound the single photon component, which is widely used in QRNG \cite{han2020practical} and quantum key distribution \cite{hwang2003quantum,lo2005decoy,wang2005beating}. However, in our protocol, we have not used the decoy state analysis in our protocol because it requires several determined intensities, which in turn requires an ideal or fully characterized intensity modulator. Since our goal is to provide a protocol that does not rely on detailed device characterization, we try to prevent considering the ideal modulator. Therefore, instead, we estimate the proportion of single photons and vacuum from the phase-randomized coherent source and consider the worst-case scenario where the multi-photon components contribute.

\hfill
\\
\noindent \textit{Statistical fluctuation}

In above analysis, we estimate the value of $C$ using asymptotic results $g'_{0}$, $g'_{1}$ and $g'_{2}$, and thus statistical fluctuations can cause errors. To account for this, we consider the experiment results $g_{e0}$, $g_{e1}$ and $g_{e2}$ obtained from $3N_{t}$ test rounds and use the Chernoff-Hoeffding tail inequality\cite{hoeffding1994probability} to obtain:

\begin{eqnarray}
	g'_{i}-\theta_{t}\leq g_{ei}\leq g'_{i}+\theta_{t},\qquad i=0,~1,~2
\end{eqnarray}
where $\theta_{t}=\sqrt{\ln(1/\varepsilon_{s})/(2N_{t})}$ with a failure probability of $\varepsilon_{s}$. Additionally, for the proportion of single photon and vacuum $\eta$, the practical proportion $\eta'$ will also suffer from statistical fluctuations, which can be bounded by
\begin{eqnarray}
	\eta-\theta\leq \eta'\leq \eta+\theta,
\end{eqnarray}
where $\theta=\theta_{t}$ for the test rounds and $\theta=\theta_{g}=\sqrt{\ln(1/\varepsilon_{s})/(2N_{g})}$ for the generation rounds. To consider the worst-case scenario caused by statistical fluctuations, we can bound the value of min-entropy and $C$ by:

\begin{eqnarray}
	&&H_{\min}(A^{n}|E^{n})_{\rho_{0}^{\otimes n}}\geq- N_{g}(\eta+\theta_{g})\log_{2}\left(  1-\frac{C}{2}\left( 1-\sqrt{1-C^{2}}\right)\right) \nonumber\\
	&&C\geq\frac{1}{\eta+\theta_{t}}\sqrt{(g_{e1}-g_{e0}-2(1-\eta)-4\theta_{t})(g_{e0}-g_{e2})},
\end{eqnarray}
with a failure probability of $6\varepsilon_{s}$. (Note that here we consider the experimental result $g_{e0}\geq (g_{e1}+g_{e2})/2$ as discussed above. If $g_{e0}\leq (g_{e1}+g_{e2})/2$, based on the symmetric concave property of the lower bound function, the lower bound of $C$ will become $C\geq\frac{1}{\eta+\theta_{t}}\sqrt{(g_{e1}-g_{e0})(g_{e0}-g_{e2}-2(1-\eta)-4\theta_{t})} $.)

To determine the final randomness rate, we use the quantum leftover hash lemma \cite{tomamichel2011leftover}. This allows Alice to extract a $\Delta$-secret random string of length $l$ through $universal_{2}$ hash function, such that:

\begin{eqnarray}
	\Delta\leq\frac{1}{2}\times2^{\sqrt{l-H_{\min}(A^{n}|E^{n})_{\rho_{0}^{\otimes n}}}}.
\end{eqnarray}
We choose the failure probability of $\Delta=\varepsilon$. Therefore, the length $l$ of the final extracted randomness bits can be determined as

\begin{eqnarray}\label{15}
	&&l\geq - N_{g}(\eta+\theta_{g})\log_{2}\left(  1-\frac{C}{2}\left( 1-\sqrt{1-C^{2}}\right)\right)-2\log_{2}\frac{1}{2\varepsilon}.
\end{eqnarray}
We select $\varepsilon_{s}=\varepsilon$, and considering the composable security, the overall failure probability is $\varepsilon_{t}=7\varepsilon$.

\subsection{Assumptions fulfillment in our implementation}

In this section, we will discuss how we can fulfill the assumptions in our implementation. The assumption (\romannumeral1) is a fundamental requirement for our protocol. It is important to note that we must have a secure location for the implementation. Fortunately, this condition is reasonable for a QRNG protocol and easily satisfied in our laboratory environment. Moving on to assumption (\romannumeral2), as discussed earlier, one of the conditions that must be met is that the encoding space is independent of the photon number space. Additionally, we also need to avoid other degrees of freedom, such as the orbital angular momentum, from carrying the modulation information, although the pulse is limited to a single photon. Fortunately, in our implementation, different phase and polarization modulations do not typically affect other properties of the input light, such as its intensity. This ensures the independence of the encoding space and the photon number space and supports our simulation of the qubit using a phase-randomized coherent source.

For assumption (\romannumeral3), it is important to consider the control of modulation noise associated with the generation state when using phase and polarization modulators that execute a unitary operator. In the case of uniform modulation fluctuations, it is possible to view every mixed state as the integration of pure states with varying fluctuations. This implies that every mixed state $ \rho_{i}$ can be expressed by

\begin{eqnarray}
	\rho_{i}&=&\int_{\varphi=0}^{2\pi}\int_{\theta=0}^{\theta'_{i}}p(\varphi)p(\theta)(\cos(\frac{\theta}{2})\ket{\omega_{i}}+\exp(i\varphi)\sin(\frac{\theta}{2})\ket{\omega_{i\perp}})(\cos(\frac{\theta}{2})\bra{\omega_{i}}+\exp(-i\varphi)\sin(\frac{\theta}{2})\bra{\omega_{i\perp}})\nonumber\\
	&=&(\frac{1}{2}+\frac{\sin(2\theta'_{i})}{4\theta'_{i}})\ketbra{\omega_{i}}+(\frac{1}{2}-\frac{\sin(2\theta'_{i})}{4\theta'_{i}})\ketbra{\omega  _{i\perp}},\qquad i=0,1,2
\end{eqnarray}
where the probability density functions $p(\varphi)$ and $p(\theta)$ satisfy the conditions $\int_{\varphi=0}^{2\pi}p(\varphi)=1$ and $\int_{\theta=0}^{\theta'_{i}}p(\theta)=1$. $\ket{\omega_{i}}$ and $\ket{\omega_{i\perp}}$ represent the eigenvectors of the state $\rho_{i}$, while $\theta'_{i}$ represents the range of fluctuations in the Bloch sphere. Considering uniform fluctuations in $p(\varphi)$ and $p(\theta)$, the length of the Bloch vector of state $\rho_{i}$ is determined by the noise range $\theta'_{i}$. Thus, in our implementation, we choose the generation state to correspond to the fewer noise point of the phase modulator in the Sagnac loop. In fact, assumption (\romannumeral3) is introduced to ensure security when considering that the noise is known to Eve. However, if we assume that the classical modulation fluctuations in the source are private, this assumption is not necessary for the security. Considering the assumption (\romannumeral4), one of the main problems that affects the modulator is charge accumulation in the birefringence modulator. However, this issue only affects modulation slower than 1 Hz \cite{wooten2000review,lunghi2015self,lin2022certified}, thus assumption (\romannumeral4) is satisfied.

It should be noted that the assumptions for the measurement devices in our protocol differ from those in measurement-device-independent quantum key distribution (MDI QKD) \cite{lo2012measurement}. In MDI QKD, measurement devices can be placed in an untrusted environment, allowing Eve to obtain all the outputs. However, since the goal of QRNG is different from that of QKD, it is reasonable to assume that the measurement is carried out in a secure environment to prevent Eve from obtaining the final random bits through public outputs and post-processing algorithms. This assumption for the measurement is also made in DI QRNGs \cite{pironio2010random,christensen2013detection,acin2016certified,bierhorst2018experimentally,liu2018device,liu2021device} and DI QKDs \cite{acin2007device,pironio2009device}. Nonetheless, in our measurement devices, Eve is allowed to preset an ancillary state which may be entangled with her states, enabling her to try to predict the outputs using this ancillary state.

\subsection{Experiment data}

\begin{figure*}[htbp]
	\centering
	\subfigure[]{\includegraphics[width=3in]{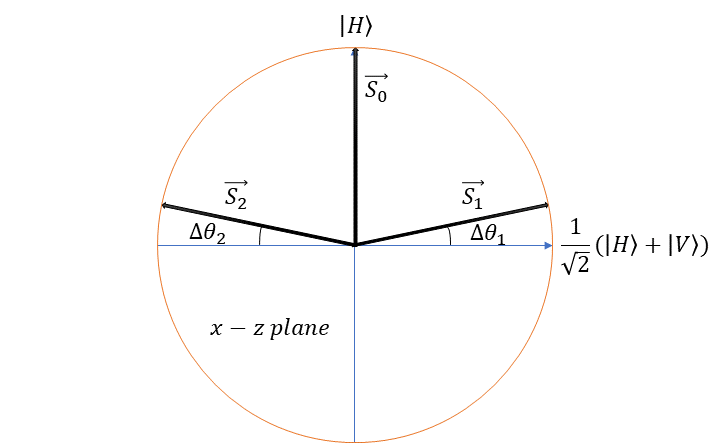}}\label{error}
	\hspace{1in} 
	\subfigure[]{\includegraphics[width=3.2in]{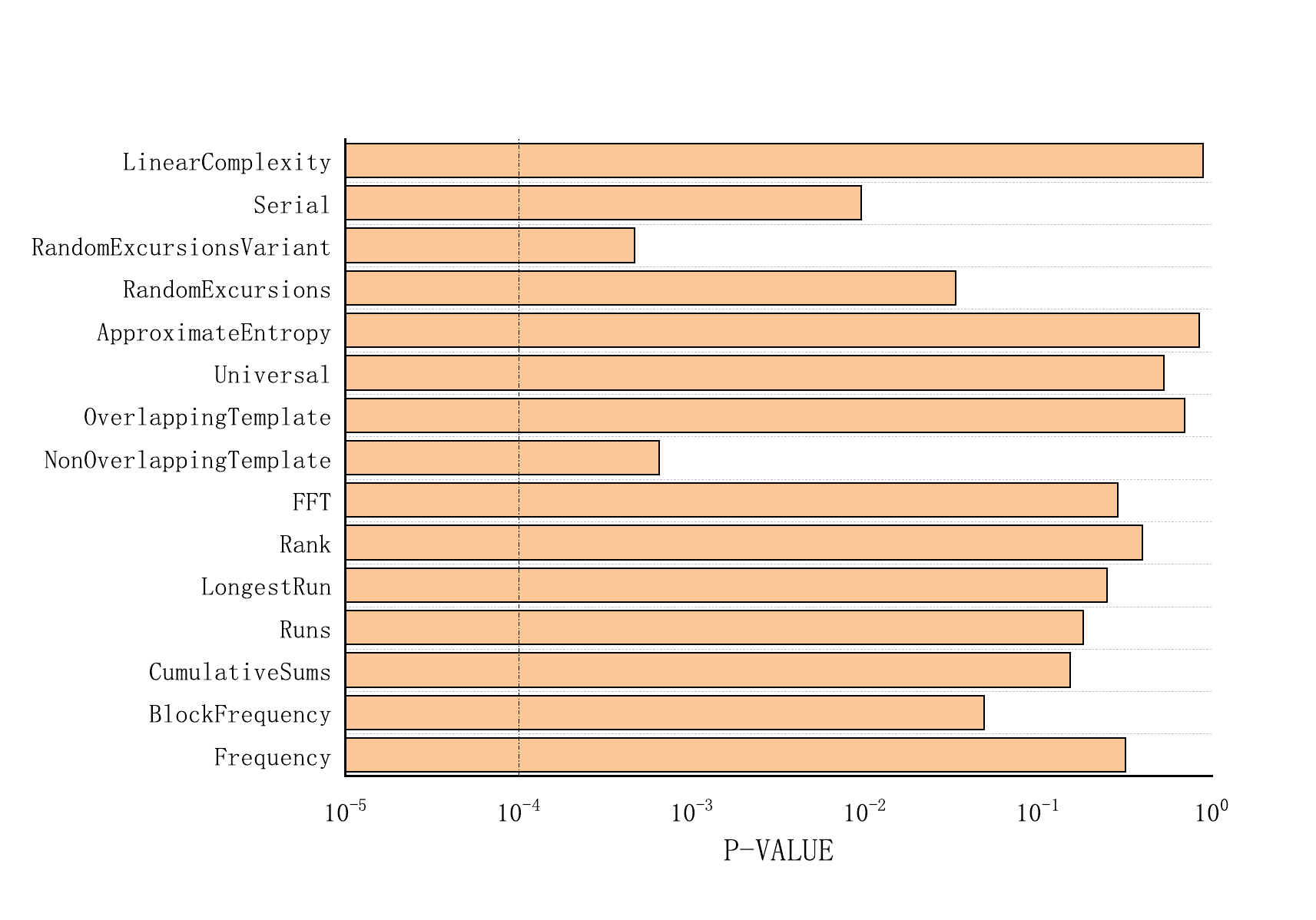}}\label{nistsfpro}
	\hspace{0.5in} 
	\subfigure[]{\includegraphics[width=3.2in]{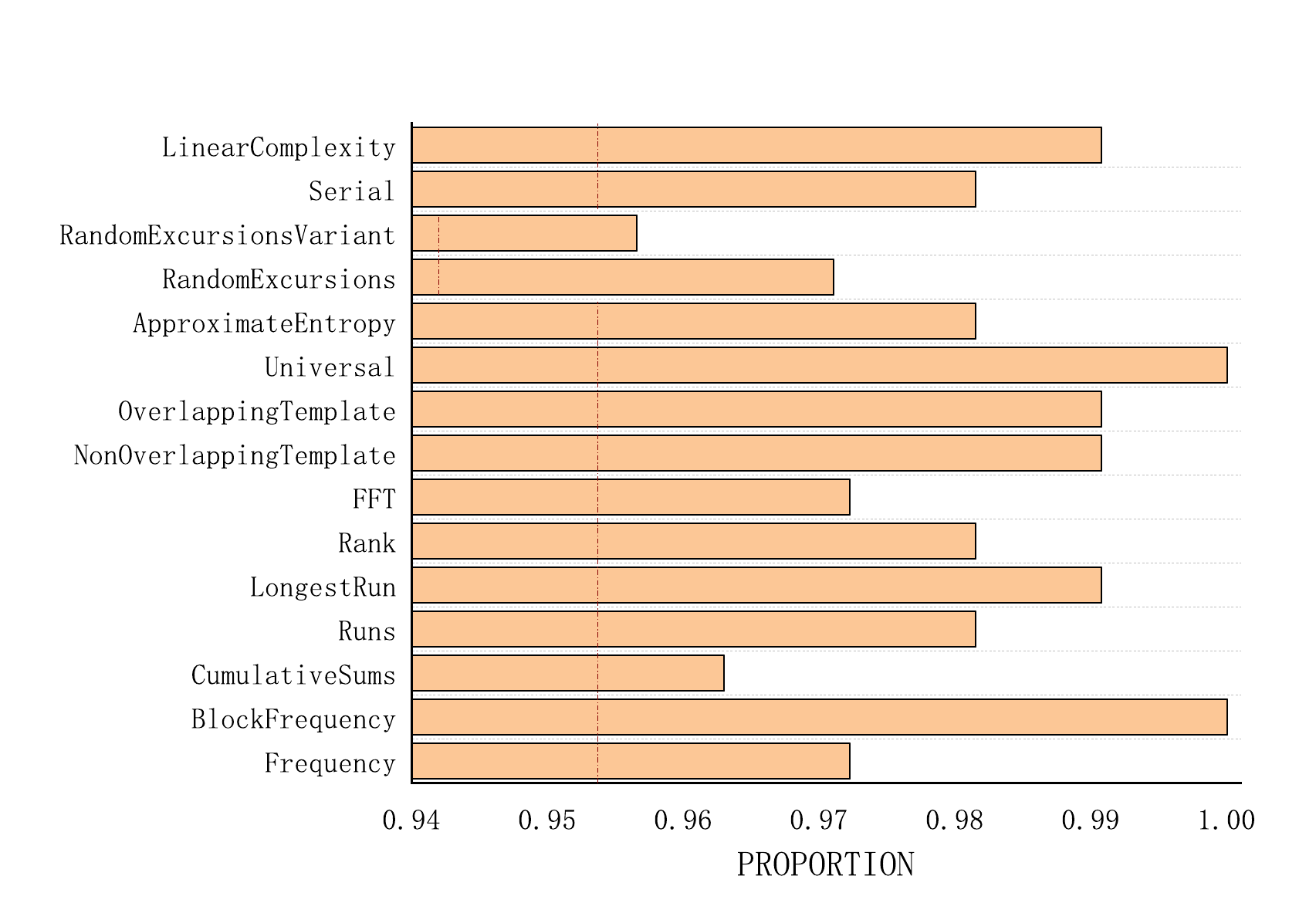}}\label{nistsfp}
	\caption{(a) shows a example of misalignment errors $\Delta\theta_{1}$ and $\Delta\theta_{2}$ for $\rho_{1}$ and $\rho_{2}$ in the Bloch sphere. (b) and (c) show the results of the NIST test with proportion and p-value. The black dotted line is the passing line. All of the test items are passed.}
	\label{fig:nistsf}
\end{figure*}

Table 1 presents the experiment data for different intensities. Specifically, we show the results for misalignment errors $\Delta\theta_{m}$ of $\frac{\pi}{14}$, $\frac{\pi}{12}$, and $\frac{\pi}{9}$. In Fig. \ref{fig:nistsf}, we show the misalignment errors $\Delta\theta_{1}$ and $\Delta\theta_{2}$ for $\rho_{1}$ and $\rho_{2}$ in the Bloch sphere. The total error satisfies $\Delta\theta_{m}=\Delta\theta_{1}+\Delta\theta_{2}$.  Here, we select the highest rate of 40.415 kbps from the experiment, corresponding to $\mu=0.58$, and generate 27 Gbit of raw data, including 270 kbit of test data. After applying the post-processing algorithm of the universal$_{2}$ hash function using the Toeplitz matrix, we obtain 108 Mbit of final randomness data. To choose the input states in the test round, we consume 35 bits to choose the position and 2 bit to choose the state for each test state, resulting in a total consumption of 10 Mbit of random numbers. To verify the statistical properties of the final data, we use the NIST SP 800-22 test suite \cite{rukhin2001statistical}. The results of the p-value and proportion in the test are shown in Fig. \ref{fig:nistsf}, and all the tests are passed.
\begin{table}[htbp]
	\setlength{\abovecaptionskip}{0cm} 
	\setlength{\belowcaptionskip}{0.2cm}
	\centering
	\caption{The experiment results of different intensities with different misalignment errors. $\mu$, mean photon number; $l$, the final extracted randomness rate. }
	$\Delta\theta_{m}=\frac{\pi}{14}$
	\\
	\begin{tabular}{p{3cm}<{\centering}|p{1.5cm}<{\centering}p{1.5cm}<{\centering}p{1.5cm}<{\centering}p{1.5cm}<{\centering}p{1.5cm}<{\centering}p{1.5cm}<{\centering}}
		\hline
		$\mu$  & 0.21 & 0.33 & 0.49 & 0.58 & 0.78 & 0.89 \\
		\hline
		$C$ &0.13572  &0.18477 & 0.224949 & 0.22938 &0.203536  & 0.155877 \\
		\hline
		$l$(bps) & 8874.4 & 22204.7 &38934.8  & 40415.4 & 26480.1 & 11425.8 \\
		\hline
	\end{tabular}\\$\Delta\theta_{m}=\frac{\pi}{12}$
	\\
	\begin{tabular}{p{3cm}<{\centering}|p{1.5cm}<{\centering}p{1.5cm}<{\centering}p{1.5cm}<{\centering}p{1.5cm}<{\centering}p{1.5cm}<{\centering}p{1.5cm}<{\centering}}
		\hline
		$\mu$  & 0.21 & 0.33 & 0.49 & 0.58 & 0.78 & 0.89 \\
		\hline
		$C $ &0.123149  &0.169014 & 0.211249 & 0.213496 &0.16917  & 0.118635 \\
		\hline
		$l$(bps) & 6607.6 & 16952.3 &32176.2  & 32505.6 & 15119.8 & 4986.6 \\
		\hline
	\end{tabular}\\$\Delta\theta_{m}=\frac{\pi}{9}$
	\\
	\begin{tabular}{p{3cm}<{\centering}|p{1.5cm}<{\centering}p{1.5cm}<{\centering}p{1.5cm}<{\centering}p{1.5cm}<{\centering}p{1.5cm}<{\centering}p{1.5cm}<{\centering}}
		\hline
		$\mu$  & 0.21 & 0.33 & 0.49 & 0.58 & 0.78 & 0.89 \\
		\hline
		$C $ &0.0989554 &0.135744 & 0.161014 & 0.158786 &0.0660784  & 0 \\
		\hline
		$l$(bps) & 3392.1 & 8726.3 &14133.8  & 13255.4 & 834.4 & 0 \\
		\hline
	\end{tabular}
\end{table}

\end{document}